\documentstyle[preprint,prl,aps]{revtex}

\begin{document}
\preprint{\vbox{\hbox{March 1995}\hbox{IFP-711-UNC}}}
\draft
\title{Dilepton Production in High Luminosity Multi-GeV Electron Scattering}
\author{\bf Paul H. Frampton and B. Charles Rasco}
\address{Institute of Field Physics, Department of Physics and Astronomy,\\
University of North Carolina, Chapel Hill, NC  27599-3255}

\maketitle

\begin{abstract}
We consider the production of a 300GeV dilepton in very intense 4GeV electron
scattering off of a lead target.
The production cross-section and angular
distribution of the resulting
muons are calculated. There occur several such events per year, and their
detection is rendered feasible by measurement of $e^+\mu^-\mu^-$ angular
correlations.
\end{abstract}
\newpage
It is generally assumed that to produce a very massive particle with a mass of,
say, 300 GeV
the best means is to employ a collider with center of mass energy well
above the particle's mass and to search for its production as an on-shell
state.
While this is generally true, we find it worthwhile to examine whether virtual
effects of the heavy particle can be detected in accelerators with
much lower energy. This is motivated in part by the temporary absence of a
10 TeV hadron or 1 TeV lepton collider. The production cross-section will be
suppressed at lower energy but this suppression can be compensated
by higher luminosity. In the present paper we consider, as an example,
the production of the dilepton (a doubly-charged gauge boson) predicted by
the 331 model\cite{331}. Our analysis is a generalization of one performed
earlier\cite{Agrawal} at high energies. First we shall evaluate the
Feynman diagrams for the process $e^{-}p \rightarrow e^{+}p \mu^{-}
\mu^{-}$ at general energies then we shall consider production
in a multi-GeV highly-luminous electron beam scattering off a stationary
nuclear target.

The basic process that we consider is $e^{-}p \rightarrow e^{+}p \mu^{-}
\mu^{-}$. The individual family lepton number violation is the
reason this signal is so clear; there is no Standard Model
background. In order to compute $e^{-}p \rightarrow e^{+}p \mu^{-} \mu^{-}$
, we must first compute $e^{-}q \rightarrow e^{+}q \mu^{-} \mu^{-}$.
This cross-section is computed from the Feynman diagrams shown
in Fig.(1), the same as were considered in reference \cite{Agrawal}. This cross
section is then folded into the proton cross section by using the EHLQ
\cite{ehlq} proton structure functions. Using the Feynman rules in
reference \cite{FrNg},
the amplitudes for $e^{-}q \rightarrow e^{+}q \mu^{-} \mu^{-}$ are

\begin{mathletters}
\label{totalamps} 
\begin{eqnarray}
Amp(a)&=& \left( \frac{g_{3L}}{\sqrt{2}} \right) e^{2}Q_{q}
\frac{(-1)}{(p_{2} - p_{4})^{2}} \frac{(-1)}{(p_{5} + p_{6})^{2} -
M_{Y}^{2} +i M_{Y}\Gamma_{Y}}\nonumber \\
& &\times \frac{1}{(p_{3}+p_{5}+p_{6})^{2}}M(a),
\label{equationa}
\end{eqnarray}
\begin{eqnarray}
Amp(b)&=&\left( \frac{g_{3L}}{\sqrt{2}} \right) e^{2}Q_{q}
\frac{(-1)}{(p_{2} - p_{4})^{2}} \frac{(-1)}{(p_{5} + p_{6})^{2} -
M_{Y}^{2} +i M_{Y}\Gamma_{Y}} \nonumber \\
& &\times \frac{1}{(-p_{1}+p_{5}+p_{6})^{2}}M(b),
\label{equationb}
\end{eqnarray}
\begin{eqnarray}
Amp(c)&=&2 \left( \frac{g_{3L}}{\sqrt{2}} \right) e^{2}Q_{q}
\frac{(-1)}{(p_{2} - p_{4})^{2}} \frac{(-1)}{(p_{1} - p_{3})^{2} -
M_{Y}^{2} +i M_{Y}\Gamma_{Y}} \nonumber\\
& &\times \frac{(-1)}{(p_{5} - p_{6})^{2} - M_{Y}^{2} +i M_{Y}
\Gamma_{Y}}M(c),
\label{equationc}
\end{eqnarray}
\begin{eqnarray}
Amp(d)&=&\left( \frac{g_{3L}}{\sqrt{2}} \right) e^{2}Q_{q}
\frac{(-1)}{(p_{2} - p_{4})^{2}} \frac{(-1)}{(p_{1} - p_{3})^{2} -
M_{Y}^{2} +i M_{Y}\Gamma_{Y}} \nonumber \\
& &\times\frac{1}{(p_{1}-p_{3}-p_{5})^{2}}M(d),
\label{equationd}
\end{eqnarray}
\begin{eqnarray}
Amp(e)&=&\left( \frac{g_{3L}}{\sqrt{2}} \right) e^{2}Q_{q}
\frac{(-1)}{(p_{2} - p_{4})^{2}} \frac{(-1)}{(p_{1} - p_{3})^{2} -
M_{Y}^{2} +i M_{Y}\Gamma_{Y}} \nonumber \\
& &\times \frac{1}{(p_{1}-p_{3}-p_{6})^{2}}M(e),
\label{equatione}
\end{eqnarray}
\end{mathletters}

with M(a), M(b), M(c), M(d) and M(e) defined as

\begin{mathletters}
\label{amplit}
\begin{eqnarray}
M(a)&=&\bar{u}(p_{4})\gamma_{\alpha}u(p_{2})\bar{u}(p_{6})\gamma_{\mu}
\gamma_{5}C\bar{u}^{T}(p_{5})\nonumber \\
& &\times v^{T}(p_{3})C\gamma^{\mu}\gamma_{5}\gamma^{\beta}
(p_{3}+p_{5}+p_{6})_{\beta}\gamma^{\alpha}u(p_{1}),
\label{equationa}
\end{eqnarray}
\begin{eqnarray}
M(b)&=&\bar{u}(p_{4})\gamma_{\alpha}u(p_{2})\bar{u}(p_{6})\gamma_{\mu}
\gamma_{5}C\bar{u}^{T}(p_{5})\nonumber \\
& &\times v^{T}(p_{3})C\gamma^{\alpha}\gamma^{\beta}
(-p_{1}+p_{5}+p_{6})_{\beta}\gamma^{\mu}\gamma_{5}u(p_{1}),
\label{equationb}
\end{eqnarray}
\begin{eqnarray}
M(c)&=&\bar{u}(p_{4})\gamma^{\alpha}u(p_{2})\bar{u}(p_{6})\gamma^{\mu}
\gamma_{5}C\bar{u}^{T}(p_{5}) v^{T}(p_{3})C\gamma^{\beta}\gamma_{5}u(p_{1})
\nonumber \\
& &\times [(p_{2}-p_{4}+p_{5}+p_{6})_{\beta}g_{\mu \alpha} +
(p_{3}-p_{5}-p_{6}-p_{1})_{\alpha}g_{\mu \beta} \nonumber \\
& &+ (p_{1}+p_{4}-p_{3}-p_{2})_{\mu}g_{\alpha \beta}],
\label{equationc}
\end{eqnarray}
\begin{eqnarray}
M(d)&=&\bar{u}(p_{4})\gamma_{\alpha}u(p_{2})\bar{u}(p_{6})\gamma^{\alpha}
\gamma^{\beta}(p_{1}-p_{3}-p_{5})_{\beta}\gamma^{\mu}\gamma_{5}C
\bar{u}^{T}(p_{5})\nonumber \\
& &\times v^{T}(p_{3})C\gamma_{\mu}\gamma_{5}u(p_{1}),
\label{equationd}
\end{eqnarray}
\begin{eqnarray}
M(e)&=&\bar{u}(p_{4})\gamma_{\alpha}u(p_{2})\bar{u}(p_{6})\gamma^{\mu}
\gamma_{5}\gamma^{\beta}(p_{1}-p_{3}-p_{6})_{\beta}\gamma^{\alpha}
C\bar{u}^{T}(p_{5})\nonumber \\
& &\times v^{T}(p_{3})C\gamma_{\mu}\gamma_{5}u(p_{1}).
\label{equatione}
\end{eqnarray}
\end{mathletters}

These amplitudes are most easily evaluated by using the method of helicity
amplitudes \cite {KleiStir}, since it is a good approximation
that all of the relevant fundemental particles are nearly massless even at
$\sqrt{s}=2.9$ GeV. The helicity amplitudes are explicitly shown in
reference \cite{Agrawal}. The $e^{-}p \rightarrow e^{+}p \mu^{-} \mu^{-}$
is computed with the EHLQ structure functions (set 1), $F_{q}(x,Q^{2})$.
The relevant cross section is

\begin{equation}
\sigma (s,M_{Y})=\int_{0}^{1}dx \sum_{q}
F_{q}(x,Q^{2})\hat{\sigma}(\hat{s}=xs,M_{Y}).
\end{equation}

Now we consider several relevant values for $\sqrt{s}$ and $M_{Y}$. We first
consider scattering off an individual proton of an electron
with beam energy on a stationary target ranging from 2 GeV up to $10^7$ GeV;
this covers everything from CEBAF(Continuous Electron Beam Accelerator
Facility) at the low end through HERA to LEPII-LHC at the high end. The
result is depicted in Fig.(2) for a dilepton mass of 300 GeV.
We see that for the cross-section
increases by some nine orders of magnitude between CEBAF
center-of-mass energies and LEPII-LHC center-of-mass energies.

A comparision of the number of events expected per year at different
colliders is as follows. At $\sqrt{s}=1790$ GeV (LEPII-LHC) equivalent to
$E_e = 1.7 \times 10^6 $ GeV, the cross-section for a proton cut-off
$p \leq 100$ MeV is 0.015pb.
The projected luminosity is $ 2 \times 10^{32} cm^{-2} s^{-1}
= 6,000 pb^{-1}y^{-1}$. This translates into 90 events/year. At $\sqrt{s}=
314$ GeV (HERA) equivalent to
$E_e = 5.2 \times 10^4 $ GeV, the cross-section
is $8 \times 10^{-6}$pb.
The luminosity is $1.6 \times 10^{31}$$cm^{-2}$$s^{-1}$
= 500 $pb^{-1}y^{-1}$. This translates into less than $10^{-3}$ events/year.
At HERA we confirm the conclusion of \cite{Agrawal} that the event-rate
is too small to have a realistic chance of discovering the dilepton.
This is because the center-of-mass energy was unfortunately chosen too low.

At $\sqrt{s}=2.9$ GeV (CEBAF) equivalent to
$E_e = 4$ GeV, the cross-section
is $9\times10^{-11}$pb.
At CEBAF the relevant process is not $e^-p\rightarrow e^+p\mu^-\mu^-$, but
 $e^-(Pb)\rightarrow e^+(Pb)\mu^-\mu^-$.
We relate these two processes by conservatively assuming an incoherent
superposition
of protons in the lead nucleus.
The
contribution of neutrons to the cross section only
increases the estimate of the total dilepton production; accurate estimates
of this correction and the effects of coherent enhancement in the nucleus
are not justified by the accuracy we employ in this paper but may become
so if unexpected events are detected.
With these approximations the projected $e^{-}p$ luminosity
 for a lead (Pb)  target is
$3.4\times10^{39}$$cm^{-2}$$s^{-1}
=1.1\times10^{11}$$pb^{-1}$$y^{-1}$. This translates into between 1 and
100  events/year depending on the value of $M_Y$.

Of the planned machines, the LEPII-LHC ( which is, presumably, at least a
decade in the
future ) would likely be the first $e^{-}N$ device which would be comparable to
CEBAF for dilepton discovery. Thus, for the remainder of this article we shall
focus specifically on the latter case. We shall consider:
(i) background estimation,
(ii) angular dependence of the muon and positron decay products, (iii) the
effects of
increasing the beam energy from 4 GeV to 8 or  10 GeV, and (iv) the
dependence of the total cross section on the dilepton mass $M_Y$.

(i) {\it Background Estimates.}

We are proposing to detect the process $e^{-}p \rightarrow e^{+}\mu^-\mu^-X$.
{}From QCD background, there will be a large number of muons arising from pion
 decay. A typical pion multiplicity at CEBAF energy (4 GeV $e^{-}$ on
nucleons) is $\sim 2$ \cite{muller} and the
cross-section is a little above 10$\mu$b.  With the design luminosity of
 $10^{39}/cm^2/s$ \cite{cebaf1} this implies on the order of $10^{17}$ pions
per year produced, about two-thirds of which
are charged according to isotopic spin invariance.  Essentially all charged
pions decay into muons.

Nevertheless, the process of interest, although it is at the level of only
 1 to 100 events per year has several specific signatures, particularly the
angular dependence of the produced
muons and positron. With a triple coincidence of $e^{+}\mu^{-}\mu^{-}$
there is a realistic chance of detecting the dilepton signal.
The key to the identification lies in the angular dependence of the products
to which we now turn.

(ii) {\it Angular Dependence.}

Let the polar angles of the positron be $(\theta,\phi)$ and those for the two
muons be $(\alpha_1,\phi_1)$ and $(\alpha_2,\phi_2)$. One azimuthal angle is
provided by the definition of a plane containing the initial beam
direction, hence there are
five independent angles describing the correlations of $e^+\mu^-\mu^-$.
Let us first focus on $\theta$.
The positron in $e^{-}p \rightarrow e^{+}\mu^{-}\mu^{-}X$ is
preferentially produced forward in the center-of-mass frame as shown in
Fig.(3). So
the first of our three coincidence triggers is on such a forward $e^{+}$
particle.

The relative angle between the two muons $\beta$ depends on the polar angles
according to:
\begin{equation}
cos\beta = sin\alpha_1sin\alpha_2cos(\phi_1-\phi_2) + cos\alpha_1cos\alpha_2
\end{equation}
Each $\mu^{-}$ is produced approximately as $1+cos^2(\alpha_i)$ in the
center-of-mass frame,
as indicated in Fig.(4), and as intuitively expected by the slowness of the
spin 1 dilepton in the center-of-mass frame at this low a beam energy.
Finally, we need to see {\bf two} $\mu^{-}$ and the dependence on $\beta$ is
dependent on the relative azimuth of the muons. We find that the muons are
preferentially emitted with $\beta < \pi/2$. This leads us to look at the
relative
angle between the positron and an emitted muon; this dependence is shown in
Fig.(5). Fig.(5) shows the probability that a muon is produced at the
corresponding $cos(\alpha_i)$ with the relative azimuth between the positron
and the muon of pi (solid line) or
a relative azimuth of zero (dashed line), assuming that the positron was
detected with $.6 < cos(\theta) < .8$ (Its most probable values).

The physical interpretation of a typical event is as follows: the
virtual dilepton is not quite at rest in the center-of-mass frame. The
dilepton
transverse momentum is approximately opposite to the positron transverse
momentum because the proton final
momentum tends to be nearly parallel to its initial momentum. When the highly
virtual dilepton decays into the two muons they carry off its momentum.
The dilepton momentum is nonrelativistic with respect to its mass, ${\bf
p}^2 \ll M_Y^2$, but this momentum is relativistic for all other particles
concerned. These calculations were all performed with $M_Y=300$ GeV.

To confirm the consistancy of Figs.(3)-(5) we have also computed the energy
distributions of the final state particles in the $e^- q$ center-of-mass
frame. If we let $\hat s$ be the squared $e^-q$ center-of-mass energy we find
the most likely
values the final state positron and quark energies are $E_{e^+} \sim
.45\sqrt{\hat s}$ and $E_q \sim .3 \sqrt{\hat s}$.
 As can be seen from Fig. (3) the
most likely angle for the $e^+$ is $cos(\theta)\sim .7$. A crude estimate of
$\hat s$ is obtainable from the center-of-mass electron  and proton
four-momenta as
$\hat s =(p_{e^-} + x p_p)^2$ where $x$ is the fraction of proton momentum
carried by the quark. If we assume the positron and the quark are emitted
near their most likely angle and near their most likely energy, independent of
$x$ this gives the ($\mu^-
\mu^-$) 3-momentum at approximately $90^o$ in the $e^- q$ CM frame. This
simple picture can help clarify the complex angular correlations presented in
Figs. (3)-(5).

This unusual decay of two like-sign muons in the same hemisphere with
transverse momentum approximately balancing that of a positron in the other
hemisphere is unique.
The realistic hope is then that this triple coincidence of $e^{+}
\mu^{-}\mu^{-}$ will be sufficiently enhancing (even to the extent
of $10^{15} \sim 10^{17}$) to compete successfully with QCD background.

(iii) {\it Effects of Increasing Beam Energy.}

If the beam energy at CEBAF were increased to 8 GeV ($\sqrt{s}=4.0$ GeV)
or 10 GeV ($\sqrt{s}=4.4$ GeV) the cross-section
increases to $2 \times 10^{-10}$pb or $3 \times 10^{-10}$pb, increasing the
number of events per year from 10 to 22 or 33  events/year respectively,
with uncertainties of an order of magnitude. At these
energies the $e^{-}p\rightarrow e^{+}p \tau^{-} \tau^{-}$ cross section
becomes kinematically possible and hence provides another detectable family
lepton
number violating process.

(iv) {\it Dilepton Mass Dependence.}

For any $\sqrt{s}\ll M_{Y}$ the total cross section goes as $M_{Y}^{-4}$.
This is easily seen as Amp(c) in equation
($\ref{totalamps}$) is suppressed because it is approximately proportional to
$M_{Y}^{-4}$ (for $\sqrt{s}\ll M_{Y}$),
due to the second dilepton propagator, while all of the other amplitudes are
approximately proportional to $M_{Y}^{-2}$,and hence Amp(c) is suppresed
by $M_{Y}^{-2}$ relative to the other amplitudes. The total cross
section goes as $M_{Y}^{-4}$.
If 2 events per year are required for detection then dileptons of up to
mass 450 GeV can be detected at current CEBAF energies.

The general conclusions of this paper are twofold: (i) to describe an exotic
event due to physics beyond the standard model, and more importantly (ii)
these events  can occur at a significant rate
in low energy high luminousity $e^-N$ colliders and
there exists the possibility of detecting new
physics at {\it e.g.} CEBAF.
Although we have, as an example, focused on a 300-450 GeV dilepton as
predicted in reference \cite{331}, our general conclusion (ii) has a wider
applicability which may merit further study.

This work was supported in part by the U.S. Department of
Energy under Grant DE-FG05-85ER-40219, Task B.

\newpage

\centerline{Figure Captions}

{\bf Figure 1.} \\
Feynman diagrams for $e^-q \rightarrow e^+\mu^-\mu^-q$ \\

{\bf Figure 2.} \\
Cross-section for $e^-p \rightarrow e^+\mu^-\mu^-p$ for
beam enegies ranging from 2GeV to $10^7$ GeV. The solid line is for transverse
momentum $p_T \geq 100$ MeV. The dashed line is for $p_T \geq 5$ GeV.\\

{\bf Figure 3.} \\
Angular distribution of $e^+$ in center-of-mass frame.\\

{\bf Figure 4.} \\
Angular distribution of $\mu^-$ in center-of-mass frame
(solid line) and $1+cos^2(\alpha)$ normalized to the real
differential cross section at $cos(\alpha) =0$ (dashed line).\\

{\bf Figure 5.} \\
Angular correlation between the positron and a $\mu^-$ with a relative
azimuth of $\Delta \phi=\pi$ (solid line) and $\Delta \phi=0$ (dashed
line);
the positron is
assumed detected with $.6 < cos(\theta) <.8$. \\

\end{document}